\documentclass[aps,prl,floatfix,twocolumn,10pt,preprintnumbers]{revtex4}

\usepackage{color,amsmath,amssymb,graphicx,latexsym,subfigure}
\usepackage{threeparttable}

\def\apj{Astrophys.\ J.\ }
\def\apjl{Astrophys.\ J.\ Lett.\ }

\usepackage{titlesec}
\titleformat{\section}[runin]{\normalfont \bfseries}{\thesection}{1em}{}

\usepackage{comment}

\newcommand{\beq}{\begin{equation}}
\newcommand{\eeq}{\end{equation}}
\newcommand{\bea}{\begin{eqnarray}}
\newcommand{\eea}{\end{eqnarray}}

\newcommand{\gsim}{\lower.7ex\hbox{$\;\stackrel{\textstyle>}{\sim}\;$}}
\newcommand{\lsim}{\lower.7ex\hbox{$\;\stackrel{\textstyle<}{\sim}\;$}}

\newcommand{\be}{\begin{equation}}
\newcommand{\ee}{\end{equation}}
\newcommand{\ba}{\begin{eqnarray}}
\newcommand{\ea}{\end{eqnarray}}

\begin{document}

\preprint{DESY\, 22-170\\\phantom{~}}
\title{Photon Ring Astrometry for Superradiant Clouds}

\author{Yifan Chen$^{a}$}
\email{yifan.chen@nbi.ku.dk}
\author{Xiao Xue$^{b,c}$}
\email{xiao.xue@desy.de}
\author{Richard Brito$^{d}$}
\email{richard.brito@tecnico.ulisboa.pt}
\author{Vitor Cardoso$^{a,d}$}
\email{vitor.cardoso@nbi.ku.dk}
\affiliation{
$^a$Niels Bohr International Academy, Niels Bohr Institute, Blegdamsvej 17, 2100 Copenhagen, Denmark\\
$^b$II. Institute of Theoretical Physics, Universit\"at Hamburg, 22761 Hamburg, Germany\\
$^c$Deutsches Elektronen-Synchrotron DESY, Notkestr. 85, 22607, Hamburg, Germany\\
$^d$CENTRA, Departamento de F\'{\i}sica, Instituto Superior T\'ecnico -- IST, Universidade de Lisboa -- UL, Avenida Rovisco Pais 1, 1049 Lisboa, Portugal
}

\begin{abstract}
Gravitational atoms produced from the superradiant extraction of rotational energy of spinning black holes can reach energy densities significantly higher than that of dark matter, turning black holes into powerful potential detectors for ultralight bosons. These structures are formed by coherently oscillating bosons, which induce oscillating metric perturbations deflecting photon geodesics passing through their interior. The deviation of nearby geodesics can be further amplified near critical bound photon orbits. We discuss the prospect of detecting this deflection using photon ring autocorrelations with the Event Horizon Telescope and its next-generation upgrade, which can probe a large unexplored region of the cloud mass parameter space when compared with previous constraints.
\end{abstract}

\date{\today}

\maketitle

\section{Introduction.}
Ultralight bosons, such as axions, dark photons, or massive tensors, are well-motivated particles beyond the standard model. They are predictions of fundamental theories with extra dimensions~\cite{Svrcek:2006yi,Abel:2008ai,Arvanitaki:2009fg,Goodsell:2009xc}, and are excellent dark matter candidates~\cite{Preskill:1982cy,Abbott:1982af,Dine:1982ah,Nelson:2011sf,Hu:2000ke} with masses above $\sim10^{-22}$ eV. 
The lightness of such fields enhances wavelike properties on astrophysical scales making their phenomenology extremely rich~\cite{Hui:2021tkt}.

An important example of this rich phenomenology occurs when the boson Compton wavelength is comparable to the gravitational radius of a rotating black hole (BH). Then, a dense bound state can be formed via superradiant extraction of the BH rotational energy~\cite{Penrose:1971uk,ZS,Brito:2015oca}. Through this process, a ``boson cloud'' can form outside the horizon, giving rise to a system also known as a gravitational atom~\cite{Detweiler:1980uk,Cardoso:2005vk,Dolan:2007mj,Arvanitaki:2009fg}.
Assuming that the BH has no external supplement of angular momentum, and considering a single unstable bound state, up to $\sim\mathcal{O}(10) \%$ of the BH mass can be transferred to the cloud~\cite{Brito:2014wla,East:2017ovw,Herdeiro:2021znw}. The local energy density of this cloud can be orders of magnitude higher than the one of virialized dark matter. Thus, the environment outside rotating BHs can be used as a powerful detector for ultralight bosons~\cite{Brito:2015oca}; this can be achieved via BH mass and spin measurements~\cite{Arvanitaki:2010sy,Arvanitaki:2014wva,Brito:2014wla,Baryakhtar:2017ngi,Brito:2017zvb,Cardoso:2018tly,Davoudiasl:2019nlo,Brito:2020lup,Stott:2020gjj,Unal:2020jiy,Saha:2022hcd}, gravitational-wave emission~\cite{Arvanitaki:2010sy,Yoshino:2012kn,Yoshino:2013ofa,Arvanitaki:2014wva,Yoshino:2015nsa,Baryakhtar:2017ngi,Brito:2017wnc,Brito:2017zvb,Isi:2018pzk,Siemonsen:2019ebd,Sun:2019mqb,Palomba:2019vxe,Brito:2020lup,Zhu:2020tht,Tsukada:2020lgt,Yuan:2021ebu,KAGRA:2021tse,Yuan:2022bem}, detection of binary BH systems~\cite{Baumann:2018vus,Hannuksela:2018izj,Baumann:2019ztm,DeLuca:2021ite,Baumann:2021fkf,Baumann:2022pkl,Cole:2022fir}, axion-induced birefringence~\cite{Chen:2019fsq,Yuan:2020xui,Chen:2021lvo,Chen:2022oad}, shadow evolution~\cite{Roy:2019esk,Creci:2020mfg,Roy:2021uye,Chen:2022nbb}, and lensing due to the extended energy distribution~\cite{Cunha:2018acu,Cunha:2019ikd,GRAVITY:2019tuf,Sengo:2022jif}. Some of these signatures benefit from the unprecedented spatial resolution achieved by the Event Horizon Telescope (EHT)~\cite{Akiyama:2019cqa,Akiyama:2019bqs,Akiyama:2019eap,Akiyama:2019fyp,EHTP,EHTM,EventHorizonTelescope:2022wkp,EventHorizonTelescope:2022wok,EventHorizonTelescope:2022exc} and are expected to improve substantially with its next generation upgrade (ngEHT)~\cite{Raymond_2021,Lngeht,Tiede:2022grp,Chael:2022meh}.

In this Letter, we introduce a new method to search for boson clouds. We propose to use the trapping properties of BHs, which are tightly connected with the existence of unstable, bound null geodesics. The existence of such geodesics is of central importance for the interpretation of BH images~\cite{1965SvA.....8..868P,1968ApJ...151..659A,Luminet:1979nyg,Campbell:1973ys,1972ApJ...173L.137C,Falcke:1999pj,MTB,Cardoso:2008bp,Cardoso:2019dte,Cardoso:2021sip,Johannsen:2010ru,Gralla:2019xty,Johnson:2019ljv,Hadar:2020fda,Tiede:2022grp}. 
For near-critical orbits, i.e., null geodesics arbitrarily close to the bound photon orbits, the motion is unbound, and photons can propagate to asymptotic observers while probing the near-horizon region. Photons from the near-critical motion circle the BH a number of times and can be used as an astrometry tool to detect oscillating lensing effects induced by real ultralight bosons.

We work in units where $\hbar = c = 1$ and adopt the metric convention $(-,+,+,+)$. Greek indices take values $(0,1,2,3)$, while spatial indices are denoted by latin letters, i.e., $x^{\mu}=(x^0, x^i)$. For clarity, we use subscripts or superscripts $V,T$ when referring to quantities related to a massive vector or tensor field, respectively.
\section{Oscillatory deflections from boson clouds.}

Consider a boson field of mass $\mu$, a BH of mass $M_{\rm BH}$, and a dimensionless angular momentum $a_J$ pointing along the $z$ axis.
When characterizing the BH-boson state, it is useful to employ a Boyer-Lindquist coordinate system. 
Ultralight bosons can bind with the BH to form a hydrogenlike gravitational atom with discrete quantum numbers~\cite{Detweiler:1980uk,Brito:2015oca,Baumann:2019eav} and a gravitational fine-structure constant $\alpha \equiv G_{N} M_{\rm BH} \mu$, where $G_{N}$ is Newton's gravitational constant. The eigenstates of the boson are characterized by an energy $\omega \sim \mu$~\cite{Detweiler:1980uk,Brito:2015oca}. When the angular velocity $\Omega$ of the BH is larger than the angular phase velocity of the bosons, i.e., $\omega/m <\Omega$, where $m$ is the azimuthal number of the boson, then a bound superradiant state is possible, which grows via extraction of rotational energy from the BH~\citep{ZS,Brito:2015oca}.

 Their energy-momentum tensor $T_{\mu\nu}$ 
 induces both static and oscillating metric perturbations in the metric. 
 For vector fields, only the traceless part of $T_{ij}^V$ oscillates. Therefore, neglecting subdominant spatial derivatives, one finds that the oscillating part of the metric perturbation is given by $H_{ij}^V \approx -  T_{ij}^{\rm osc} / (2 \mu^2 m_{\rm pl}^2)$,  where $m_{\rm pl} \equiv 1/\sqrt{8\pi G_{N}}$ 
 is the reduced Planck mass. On the other hand, massive tensors behave as a localized  effective strain $H_{\mu\nu}^T$ proportional to its field value and a dimensionless coupling constant $\alpha_h$ characterizing the strength of the interaction between the massive tensor and photons~\cite{Armaleo:2020yml}\footnote{Previous constraints assume either couplings to fermions ~\cite{Hohmann:2017uxe,Talmadge:1988qz,Sereno:2006mw,Armaleo:2020yml} or massive tensors as the dominant dark matter component~\cite{Armaleo:2019gil,Armaleo:2020yml,Sun:2021yra,Unal:2022ooa}.}.
We introduce dimensionless parameters $\epsilon_V \equiv  (\Psi_0^{V}/m_{\rm pl})^2/2$ and $\epsilon_T \equiv \alpha_h \Psi_0^T/m_{\rm pl}$ to characterize the  perturbative strain amplitude 
generated from a vector and a tensor cloud, respectively, where $\Psi_0^V$ and $\Psi_0^T$ are their corresponding maximal field value. See the Supplemental Material~\cite{supp} for details.

Considering metric perturbations $H_{\mu\nu} \equiv \epsilon h_{\mu\nu}$ expanded around a Kerr background $g_{\mu\nu}^{K}$, we have $g_{\mu\nu} \simeq g_{\mu\nu}^{K} + \epsilon h_{\mu\nu}$, where $\epsilon  \ll 1$ controls the amplitude of the perturbations. The photon geodesics in this perturbed metric can be similarly expanded, $x^\mu \simeq x_{(0)}^\mu + \epsilon x_{(1)}^\mu$, where $x_{(1)}^\mu$ satisfies
\be
\frac{d^{2} x^{\mu}_{(1)}}{d \lambda^{2}} = -2 \Gamma_{\alpha \beta}^{{K} \mu} \frac{d x^{\alpha}_{(0)}}{d \lambda} \frac{d x^{\beta}_{(1)}}{d \lambda} - \Gamma_{\alpha \beta}^{(1)\mu} \frac{d x^{\alpha}_{(0)}}{d \lambda} \frac{d x^{\beta}_{(0)}}{d \lambda}. \label{eq:geodp}
\ee
 Here $x_{(0)}^\mu$ and $\Gamma_{\alpha \beta}^{{K}\mu}$ are the unperturbed null geodesics and the connection of the Kerr background, respectively, and $\lambda$ is an affine parameter. 
 $\Gamma_{\alpha \beta}^{(1)\mu}$ is the perturbed connection generated from bosonic clouds.

We use a backward ray tracing code on a Kerr background, \texttt{KGEO}~\cite{Chael_kgeo_2022}, to compute $x_{(0)}^\mu$ from a faraway camera toward the BH  employing the integral method~\cite{Carter:1968rr,Bardeen:1973tla,Gralla:2019drh,Gralla:2019ceu}.
 {We consider an faraway observer at an inclination angle $\theta_{o} = 17^\circ$ from the black hole with dimensionless spin $a_J = 0.94$, consistent with the EHT observations of M87$^\star$~\cite{Akiyama:2019cqa}. 
We calculate the geodesics  starting at different pixels on the observer plane and keep only those that hit the BH equatorial plane at least two times. The initial ($\lambda = 0$) components of photon geodesics are fixed to be $r(\lambda=0)=10^3r_g, t(\lambda=0)=0$, without loss of generality,  where $r_g \equiv G_{N} M_{\rm BH}$ is the gravitational radius. The plasma medium effects, such as absorption and scattering effects, are neglected since the accretion flow is considered to be optically thin.}

\begin{figure}[t]
    \centering
    \includegraphics[width=0.45\textwidth]{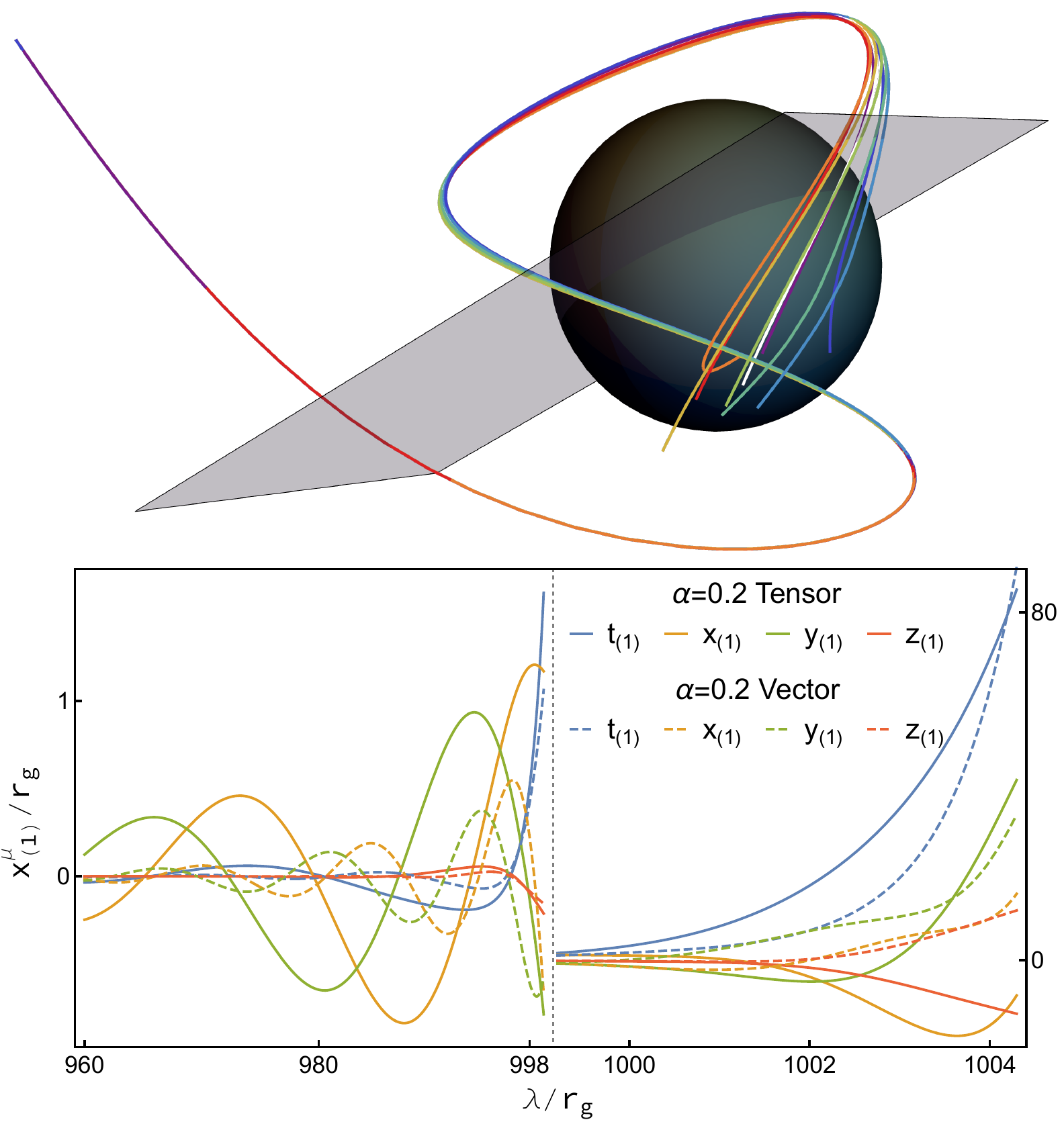}
    \caption{
    Examples of photon geodesics connecting a faraway observer from a Kerr BH surrounded by a bosonic cloud, that execute multiple orbits intersecting the equatorial plane (gray plane in top panel). 
    Top panel: white line shows the unperturbed geodesic, while different colors show perturbed geodesics $x_{(0)}^i + \epsilon_T\, x_{(1)}^i$ at different oscillation phases of bosons.
    Bottom panel: deviation of the geodesics in Kerr-Schild coordinates in terms of the affine parameter $\lambda$. 
    The dashed vertical line corresponds to the point where the unperturbed orbit $x_{(0)}^\mu$ first crosses the equatorial plane.}
    \label{fig:geo}
\end{figure}

An example of a near-critical unperturbed geodesic $x_{(0)}^{i}$ is shown in the top panel of Fig.\,\ref{fig:geo} using a white line that, due to strong lensing effects, crosses the gray equatorial plane of the Kerr BH several times. We also show samples of the perturbed spatial photon geodesics $x_{(0)}^i + \epsilon_T\, x_{(1)}^i$ for 
{eight evenly spaced initial phases to account for the full oscillation behavior of the cloud.}
As expected, deviations start to be apparent after entering the photon ring orbit.

The deviation $x_{(1)}^\mu$ is shown in the bottom panel of Fig.\,\ref{fig:geo} for a massive tensor and vector cloud in the ground state with $\alpha = 0.2$.
To avoid (unphysical) singularities at the poles, we show $x_{(1)}^\mu$ using Cartesian Kerr-Schild coordinates $(t, x, y, z)$, in units of the gravitational radius $r_g$.
The evolution of $x_{(1)}^\mu$ as a function of the affine parameter can be divided in two stages. The first stage (bottom-left panel) is dominated by perturbative deviations from cloud-induced metric fluctuations far away from the BH. In the second stage, when the orbit is close to the photon ring, deviations are exponentially amplified, as seen on the bottom right panel of Fig.\,\ref{fig:geo}. 

The second term on the right-hand side of Eq.\,(\ref{eq:geodp}) dominates during the first stage, which is especially important for photons propagating at the typical radial scale of the cloud, i.e., $r_g/\alpha^2$, where our Newtonian approximation for the cloud profile, {as discussed in the Supplemental Material~\cite{supp}}, is reasonable. 
The oscillation of the boson fields leave imprints on $x_{(1)}^\mu$, with the frequency of the oscillations for vector clouds being twice the one of the tensor cloud case. 
The exponential growth of $x_{(1)}^\mu$ starts when $x_{(0)}^\mu$ enters a nearly bound orbit where the first term in Eq.\,(\ref{eq:geodp}) becomes important. This growth is caused by the instability of the photon ring orbit~\cite{1965SvA.....8..868P,1968ApJ...151..659A,Luminet:1979nyg,Campbell:1973ys,1972ApJ...173L.137C,Falcke:1999pj,MTB,Cardoso:2008bp,Johnson:2019ljv,Gralla:2019drh}.
More precisely, a face-on observer would see a deviation increase by a factor of $\sim20$ between two sequential crossings of the BH equatorial plane.

\section{Astrometry with photon ring autocorrelations.}
The next question concerns the observability of such an effect. 
One sensitive and realistic probe is to use photon ring autocorrelations as proposed in Ref.~\cite{Hadar:2020fda}, which is an especially useful observable for BHs observed nearly face on, such as M87$^\star$ and Sgr\,A$^\star$.

The strong gravity region outside Kerr BHs is responsible for the existence of bound null geodesics and for near-critical trajectories {where photons propagate around the BH multiple times before reaching a narrow ring region on the observer plane.
These strongly lensed photons form the photon ring with a locally enhanced intensity~\cite{Johannsen:2010ru,Gralla:2019xty,Johnson:2019ljv}. 
One way to observe the strongly lensed photons, without the need of significant improvements of the current spatial resolution, is to exploit the time variability of the emission from the accretion flow, for example. Indeed, intensity fluctuations due to the flow's turbulent nature were observed by the EHT~\cite{EventHorizonTelescope:2022ago,EventHorizonTelescope:2022exc}.
For each emission point, there are multiple geodesics connecting it to the observer's image plane that differ by the number $k$ of half orbits around the BH. In the image plane, we can use polar coordinates $(\rho, \varphi)$, where $\rho$ is the radial distance from the BH and $\varphi$ is the polar angle. Photons with common origin but different $k$ reach the photon ring separated in time $t$ and angle $\varphi$, rendering the two-point intensity fluctuation correlation~\cite{Hadar:2020fda}
\begin{equation}
    \label{eq:intensity}
    \mathcal{C}(T, \Phi)\equiv  \left\langle\Delta I(t, \varphi) \Delta I\left(t\!+\!T, \varphi\!+\!\Phi\right)\right\rangle,
\end{equation}
to have peaks at $T \approx N \tau_0, \Phi \approx N \delta_0$ for a background Kerr BH, where $\Delta I$ is the intensity fluctuation after an integration in the radial direction on the photon ring, $N \equiv k - k^\prime$ is the difference between a photon pair executing $k$ and $k^\prime$ half orbits, and $\{\tau_0,\delta_0\}$ are the critical parameters characterizing the time delay and azimuthal lapse of the bound photon orbits, respectively~\cite{Gralla:2019drh}. This is a universal prediction dependent only on the space-time near the photon shell, especially for optically and geometrically thin emission flows~\footnote{The EHT observation of M87$^\star$~\cite{EHTP,EHTM} and SgrA$^\star$~\cite{EventHorizonTelescope:2022wkp,EventHorizonTelescope:2022wok,EventHorizonTelescope:2022exc,Wielgus:2022heh} favors a magnetically arrested disk model, which is geometrically thin in the inner region~\cite{Igumenshchev:2003rt, Narayan:2003by, McKinney:2012vh, Tchekhovskoy2015}.}.}

In the presence of boson clouds, small oscillations around the unperturbed geodesics grow exponentially close to the photon ring orbit, {leading to a periodic shift in the peak positions, $T\approx N \tau_0+\Delta T^N$ and $\Phi\approx N \delta_0+\Delta \Phi^N$. For photons emitted from a geometrically thin disk at the equatorial plane}, the deviations can be calculated using
\be\begin{split}
 \Delta T^N/\epsilon =& x^{t}_{(1)} (\lambda_{N}) - x^{t}_{(1)} (\lambda_{0}),\\ \Delta \Phi^N/\epsilon =& x^{\phi}_{(1)} (\lambda_{N}) - x^{\phi}_{(1)} (\lambda_{0}),\label{DeltaTPhi}\end{split}\ee
  where $\lambda_N$ represents the affine parameter at which the perturbed geodesics crosses the equatorial plane, i.e., $z_{(0)}+\epsilon\,z_{(1)} = 0$, for the $N+1$ times. Since $x^\mu_{1}$ grows exponentially after $\lambda_0$, one can safely neglect $x^{\mu}_{(1)} (\lambda_0)$ in Eq.\,(\ref{DeltaTPhi}). {Notice that one can get an equivalent gauge-invariant description of the geodesics deflection and shifts on the image plane using the deviation of the conserved quantities in the Kerr space-time instead.}

{The fundamental resolution of the autocorrelation is limited by the intrinsic correlation length of the source, which can be inferred from general relativistic magnetohydrodynamic simulations~\cite{Hadar:2020fda}. 
Taking $\ell_\phi \approx 4.3^\circ$  and $\ell_t \approx 3\, r_g$ as correlation lengths for $\phi$ and $t$, respectively~\cite{Hadar:2020fda}, $\Delta T^N/ \ell_t$ is typically smaller than $\Delta \Phi^N/\ell_\phi$. We thus focus on $\Delta \Phi^N$, fitting it to be $\Delta \Phi^N = \Phi_0^N \cos\left(\omega t + \delta\right)$ for  $N=1$ and $2$, where
$\Phi_0^N$ is the amplitude and $\delta$ is the relative oscillation phase.} The phase $\delta$ is always well fit by the sum of  $2\varphi$ and a small deviation, representing an $m = 2$ mode of metric perturbation and a small inclination angle.

\begin{figure}[htb]
\includegraphics[width=0.45\textwidth]{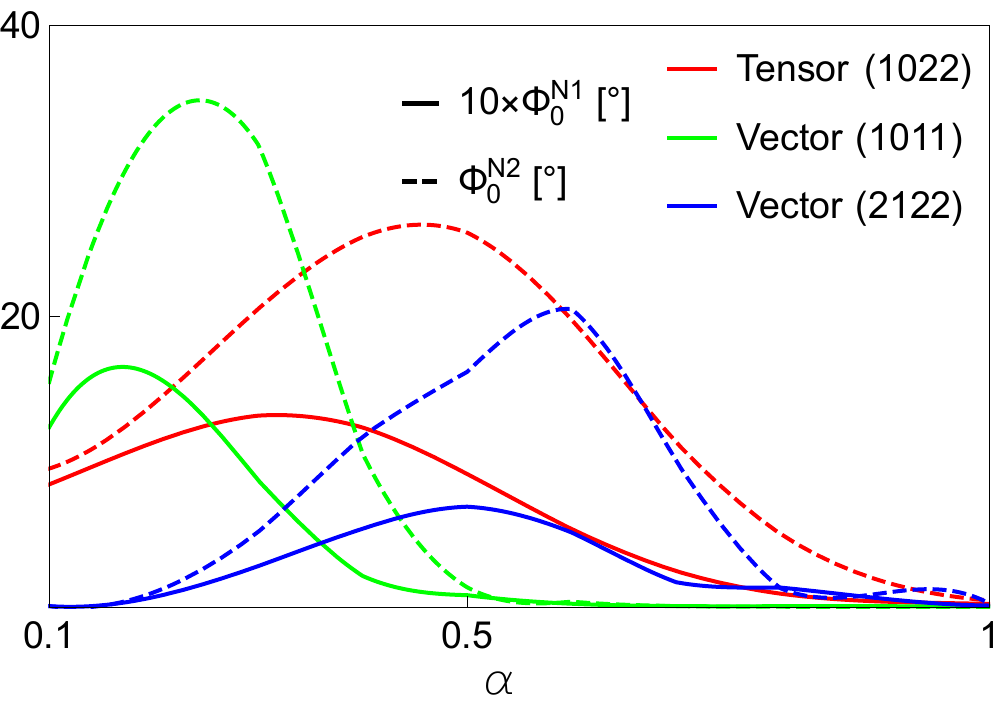}
    \caption{Oscillation amplitude $\Phi_0$ of the azimuthal lapse at $\varphi \approx 0.7\pi$ on the critical curve
    at different values of $\alpha$ with $a_J = 0.94$ , $\theta_{o} = 17^\circ$, and $\epsilon = 10^{-3}$. 
    }
    \label{fig:ampalpha}
\end{figure}
The spatially varying $\Phi_0^N$, as shown in the Supplemental Material~\cite{supp}, would require weighting the intensity fluctuation spectrum as a function of the photon ring region, which is beyond the scope of this study. We will instead estimate the detectability of the signal using a specific point on the critical curve with the largest amplitude $\varphi \approx 0.7\pi$, which is also the point with the most significant intensity contributed by the lensed photon~\cite{Chen:2022nbb}. We show $\Phi_0^N$ at this point as a function of $\alpha$ for the ground-state tensor and vector cloud in Fig.\,\ref{fig:ampalpha}.
As expected, $\Phi_0^{N2}$ is typically larger due to the exponential sequential growth.
For completeness, we also consider the higher vector field mode $(2122)$ (this notation is explained in the Supplemental Material~\cite{supp}), which has a smaller maximum value for $\Phi_0^N$ when compared to the ground-state mode due to the $\sin^2 \theta$ suppression for geodesics reaching a nearly face-on observer.

\section{Prospective constraints.}
{In a realistic setup, the finite spatial resolution of EHT or ngEHT leads to a Gaussian smearing of the correlation peak in the $(T, \Phi)$ plane. For ngEHT, assuming an $\sim10\mu$as spatial resolution~\cite{Chael:2021rjo,Lico:2023mus}, the $1\sigma$ width of the Gaussian packet in the $\Phi$ direction is around $10^\circ$ for both M87$^\star$ and Sgr A$^\star$. 
On the other hand, in the presence of a boson cloud, there will be periodic oscillations of the central position of the Gaussian packet, causing a broadening of its width that becomes especially evident when the oscillation amplitude $\Phi_0^N$ is comparable or larger than the ``original'' width without oscillations.
Thus, if a correlation peak predicted assuming a vacuum Kerr BH is observed, one can impose the conservative limit that $2 \Phi_0^N < 10^\circ$. Notice that such a conservative estimate does not exploit the full oscillatory features. Once a detection of a correlation peak has been made, one can also check whether an oscillation in the center of the Gaussian packet provides a better description than a stationary peak, which can potentially give a sensitivity beyond the one limited by the Gaussian smearing. With less than $1$ order-of-magnitude-longer observation times or a slightly improved spatial resolution, one could, in principle, resolve the intrinsic correlation of the plasma with $\ell_\phi \approx 4.3^\circ$. We thus use this limit as an optimistic criterion for our prospective constraints.}

\begin{figure}[htb]
    \centering
    \includegraphics[width=0.45\textwidth]{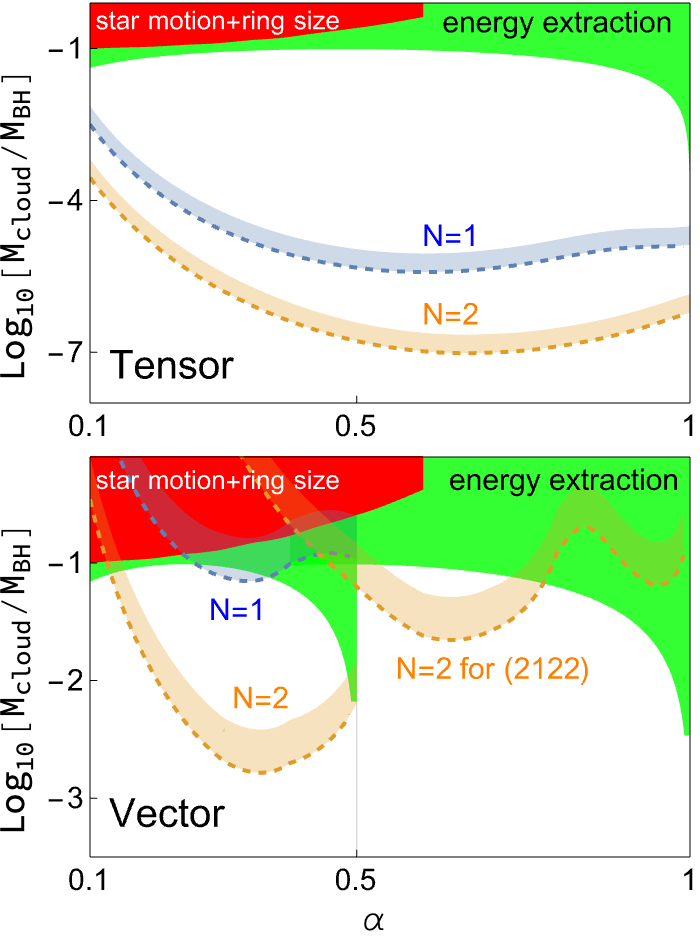}
    \caption{Prospects for constraints on the mass of a vector cloud or a tensor cloud with $\alpha_h = 1$ and a BH dimensionless spin $a_J = 0.94$. The red region comes from constraints from stellar motion jointly with ring size measurements using the current EHT with 10\% error bars~\cite{Sengo:2022jif}, and the green region is the theoretical bound on the maximum superradiant extraction for the considered modes~\cite{Herdeiro:2021znw}. {The constraint bands show limits ranging from a conservative criterium based on ngEHT's spatial resolution to an optimistic criterion based on the intrinsic azimuthal correlation length of the accretion flow.}}
    \label{fig:mfc}
\end{figure}

The oscillation amplitude $\Phi_0^N$  shown in Fig.\,\ref{fig:ampalpha} is proportional to $\epsilon_T$ or $\epsilon_V$, which in turn is related to the total mass in the cloud as discussed in the Supplemental Material~\cite{supp}. {The results are summarized in Fig.\,\ref{fig:mfc}, where the dashed lines show the optimistic limit based on the intrinsic correlation length while the bands show conservative constraints that could be obtained immediately after a detection of the corresponding autocorrelation peak.}
Here we take $\alpha_h = 1$ for massive tensors and limit the $\alpha$ range for the vector mode $(1011)$ to be below $0.5$, which is the limit imposed by assuming that the ground-state cloud formed due to superradiance.

For comparison, we also show constraints taken from Ref.~\cite{Sengo:2022jif} based on joining EHT observations of M87$^*$'s photon ring size with measurements of the BH mass-to-distance ratio obtained from stellar (or gas) dynamics. Those constraints correspond to the red region labeled ``star motion + ring size''. Furthermore, there is a theoretical bound on the maximum mass a cloud can reach from extracting BH rotation energy through superradiance~\cite{Herdeiro:2021znw}, assuming that there is no angular momentum supplement to the BH, {which 
could be breached due to a potential accretion process~\cite{Brito:2014wla,Hui:2022sri}}

The photon ring autocorrelations with the $N=1$ subring could already go far beyond previous limits for a tensor cloud, while a detection of the $N=2$ subring can probe both vector and tensor clouds sensitively. In the Supplemental Material we show similar results for different BH spins~\cite{supp}. In general, we find qualitatively similar results for smaller spins, but with smaller oscillations due to the fact that the photon ring radius gets larger for smaller spins. However, even for $a_J=0.5$, the $N=2$ subring can always constrain interesting regions of the parameter space.

\section{Discussion.}
Metric perturbations induced by superradiant clouds lead to oscillatory deflections of photon geodesics.
Because of strong gravitational effects, searching for this effect benefits from two aspects compared to previous astrometry searches for ultralight bosonic dark matter~\cite{Khmelnitsky:2013lxt,Porayko:2018sfa,Boskovic:2018rub,Guo:2019qgs,Nomura:2019cvc,Armaleo:2020yml,PPTA:2021uzb,Sun:2021yra,Unal:2022ooa,Yuan:2022nmu}: 
(i) Superradiant clouds can reach large field values outside a BH, and (ii) any deviation of the vacuum geodesics grows exponentially close to the photon ring orbit. 
Importantly, this method relies solely on observations of the photon ring. With longer integration times, the EHT could already detect the $N=1$ subring~\cite{Hadar:2020fda}, while the expected improvements in baseline coverage, spatial resolution, dynamic ranges, and {multifrequency observations of the ngEHT could help shorten the integration time}.

Another observable that we did not consider is the time delay that can be seen in Fig.\,\ref{fig:geo}, which is also the dominant observation channel for a scalar cloud. The prospects to detect such time delay are less interesting due to the limited time resolution of the photon ring autocorrelation. However, in the presence of an emission source with shorter time correlation lengths, such as a hot spot~\cite{Moriyama:2015zfa,Moriyama:2019mhz,Wong:2020ziu,Chesler:2020gtw,Gussmann:2021mjj,Andrianov:2022snn} or a nearby pulsar~\cite{Kocherlakota:2017hkn,Kimpson:2020hny,Kimpson:2020uje,Wu:2022maa,Ben-Salem:2022txj}, echoes from those sources could be used as astrometry with a dramatically enhanced time resolution similar to pulsar timing array searches.

Our discussion can also be applied to other profiles of ultralight bosons, such as a soliton core dark matter, whose typical field value is $\Psi_{\rm DM} \approx 1.4\times 10^{11}$ GeV ($\epsilon_V \sim 10^{-15}$ and $\epsilon_T \sim 10^{-7} \alpha_h$) and radius $r_{c} \simeq 40$pc for $\mu = 10^{-21}$ eV~\cite{Schive:2014dra}. Using a flat background approximation, the propagation of light in the soliton core leads to a spatial deviation $\sim \epsilon\, r_{c} \simeq 10\,\alpha_h r_g$ before entering the photon ring orbit of Sgr\,A$^\star$. Considering the amplification process in the photon ring, we estimate to be able to constrain $\alpha_h$ to be less than $10^{-4}$ using the $N=1$ subring and an order-of-magnitude better for $N=2$. These bounds are comparable with those obtained with pulsar timing arrays or planetary motion~\cite{Armaleo:2020yml}.

Finally, we should note that the oscillating cloud will also cause emission of gravitational waves (GWs). Assuming M87$^\star$ and $\alpha$ that maximizes the constraints in Fig.~\ref{fig:mfc}, the half-life~\footnote{Notice that the decay due to GW emission goes as $\propto 1/(1+t/\tau_{\rm GW})$ with $\tau_{\rm GW}$ the half-life~\cite{Brito:2014wla,Brito:2017zvb}.} of the cloud due to GW emission will typically be of order $\sim10^{4}$ years~\cite{Baryakhtar:2017ngi,Brito:2020lup}, which could potentially be significantly longer if the fields self-interact~\cite{Yoshino:2012kn,Fukuda:2019ewf,Baryakhtar:2020gao,East:2022ppo,Omiya:2022gwu}. This also opens the potential to joint detections with LISA, which could detect GWs emitted by boson clouds for $\mu\gtrsim 10^{-19}$ eV~\cite{Brito:2015oca}.

\hspace{3mm}

\begin{acknowledgments}
This work is supported by the Villum Investigator program supported by the VILLUM Foundation (Grant No. VIL37766) and the DNRF Chair program (Grant No. DNRF162) by the Danish National Research Foundation, and under the European
Union’s H2020 ERC Advanced Grant “Black holes: gravitational engines of discovery” Grant Agreement No. Gravitas–101052587, and by Fundação para a Ciência e Tecnologia I.P, Portugal under Project No. 2022.01324.PTDC.
X.X. is supported by  Deutsche Forschungsgemeinschaft under Germany’s Excellence Strategy EXC2121 “Quantum Universe” Grant No. 390833306. The views and opinions expressed are those of the author only and do not necessarily reflect those of the European Union or the European Research Council. Neither the European Union nor the granting authority can be held responsible for them.
R.B. acknowledges financial support provided by FCT, under the Scientific Employment Stimulus -- Individual Call -- Grant No. 2020.00470.CEECIND.
\end{acknowledgments}

\pagebreak
\widetext
\begin{center}
\textbf{\large Supplemental Materials: Photon Ring Astrometry for Superradiant Clouds}
\end{center}
\setcounter{equation}{0}
\setcounter{figure}{0}
\setcounter{table}{0}
\makeatletter
\renewcommand{\theequation}{S\arabic{equation}}
\renewcommand{\thefigure}{S\arabic{figure}}
\renewcommand{\bibnumfmt}[1]{[#1]}
\renewcommand{\citenumfont}[1]{#1}

\hspace{5mm}


\begin{center}
\textbf{I: Gravitational atom states}
\end{center}

For concreteness, we focus on superradiant ground states for scalars, vectors and massive tensors in the Newtonian limit.
The wavefunctions for scalars and vectors are~\cite{Detweiler:1980uk,Brito:2015oca,Baryakhtar:2017ngi}
\be
\begin{split} a^{211} &= \Psi_0^S R_\alpha^{21} \cos \left(\mu t - \phi\right) \sin \theta,\\  A_i^{1011} &= \Psi_{0}^V R_\alpha^{10} \left( \cos \left(\mu t\right), \sin \left(\mu t\right) , 0 \right),    \end{split}\label{GAWF} 
\ee
respectively, where $a$ and $A_i$ represent a scalar field and the Cartesian components of the vector field in the unitary gauge, respectively. We neglect $A_0$ due to its suppression in a small-$\alpha$ expansion. Bound states are characterized by a set of quantum numbers: $(n\ell m) = (211)$ and $(n\ell jm) = (1011)$ describe a scalar and a vector state, respectively, where $n, \ell, j, m$ are the  principal, orbital angular momentum, total angular momentum and azimuthal number (we use the notation adopted in Ref.~\cite{Baumann:2019eav} where $n\geq \ell+1$).
The constant amplitude $\Psi_0$ is used to parameterize the peak value of the wavefunction, where we set the maximal value of the hydrogenic radial wavefunctions to 1, i.e., $R_\alpha^{10} = {\rm Exp} [-\alpha^2 r/r_g]$ and $R_\alpha^{21} = {\rm Exp} [1-\alpha^2 r/(2 r_g)]\, \alpha^2 r/(2 r_g)$. From Eq.\,(\ref{GAWF}), their corresponding energy-momentum tensor is
\be\label{eq:Tmunu}
\begin{split}  T_{\mu\nu}^S \approx & \frac{1}{2} \left(\mu \Psi_0^{S} R_\alpha^{21} \sin \theta \right)^2 \times \textrm{diag} \left(1,c_S,c_S,c_S\right),\\
  T_{\mu\nu}^V \approx & \left(\mu \Psi_0^V R_\alpha^{10} \right)^2\times
 \left(\begin{array}{cccc}
1 & 0  & 0 & 0 \\
0 &  c_V &   s_V &  0 \\
0 &   s_V  & -c_V &  0 \\
0 &  0 &  0 & 0
\end{array}\right),
\end{split}
\ee
where we took the Newtonian limit and the leading order in an expansion at small $\alpha$, equivalent to neglecting all the spatial derivatives. We defined $c_S\equiv-\cos \left(2\mu t - 2 \phi\right), c_V \equiv \cos \left(2\mu t\right)$ and $s_V \equiv \sin \left(2\mu t\right)$.

 The energy-momentum tensors in Eq.~\eqref{eq:Tmunu}  induce both static and oscillating metric perturbations in the metric.  For scalar fields and in the Newtonian approximation that we are considering, only the trace part of $T_{ij}^S$ oscillates. By solving the linearized Einstein equations one finds that this leads to oscillations in both the $00$ and $\delta_{ij}$ components of the metric perturbation~\cite{Khmelnitsky:2013lxt}. Instead, for vector fields, only the traceless part of $T_{ij}^V$ oscillates in Eq.~\eqref{eq:Tmunu}. Neglecting subdominant spatial derivatives, one gets the oscillating part of the metric perturbation by $H_{ij}^V \approx -  T_{ij}^{\rm osc} / (2 \mu^2 m_{\rm pl}^2)$,  where $m_{\rm pl} \equiv 1/\sqrt{8\pi G_{N}} \simeq 2.435 \times 10^{18}$ GeV 
 is the reduced Planck mass.

On the other hand, superradiant massive tensor fields behave as a localized gravitational wave~\cite{Brito:2013wya,Aoki:2017ixz,Brito:2020lup,Jain:2021pnk} that can be characterized by the effective strain
\be H^{T}_{ij} = \frac{\alpha_h \Psi_0^T R_\alpha^{10}}{m_{\rm pl}}
  \left(\begin{array}{ccc}
 \cos \left(\mu t\right) &  \sin\left(\mu t\right) & 0 \\
 \sin \left(\mu t\right)  & - \cos \left(\mu t\right) & 0 \\
 0 & 0 & 0
 \end{array}\right),\label{hMT}
 \ee
for the hydrogenic ground state $(n\ell jm) = (1022)$, linearly dependent on the field value normalization $\Psi_0^T$. As explained in the main text, here $\alpha_h$ is a dimensionless coupling constant characterizing the strength of the interaction between the massive tensor and photons~\cite{Armaleo:2020yml}. {Constraints on $\alpha_h$ that assume couplings to fermions have been imposed using the Cassini spacecraft~\cite{Hohmann:2017uxe} and planetary dynamics~\cite{Talmadge:1988qz,Sereno:2006mw,Armaleo:2020yml}, whereas constraints based on couplings to photons, but assuming the massive tensor to be the main component of dark matter, have been imposed using binary pulsar observations~\cite{Armaleo:2019gil} and pulsar timing array observations~\cite{Armaleo:2020yml,Sun:2021yra,Unal:2022ooa}. In the main text we keep $\alpha_h$ generic since we do not need to assume that the massive tensor is dark matter or that it couples to fermions.}

Comparing the oscillating metric perturbation from a vector cloud with Eq.\,(\ref{hMT}), there is a simple mapping between the vector and tensor cloud case: $\epsilon^V \equiv  (\Psi_0^{V}/m_{\rm pl})^2/2 \rightarrow \epsilon^T \equiv \alpha_h \Psi_0^T/m_{\rm pl}$, 
  $(R_\alpha^{10})^2 \rightarrow R_\alpha^{10}$, and $2\mu t \rightarrow \mu t$, where ${\epsilon^{V/T}}$ is used to characterize the effective strain amplitude coupled to electromagnetic fields.

For a scalar cloud, if we ignore the sub-leading spatial derivatives, the metric perturbations in the trace part only cause a time delay of propagating photons~\cite{Khmelnitsky:2013lxt}.
{More explicitly, the perturbed connections are non-zero only for $\Gamma^{(1)t}_{ii}, \Gamma^{(1)i}_{it}$ and $\Gamma^{(1)i}_{ti}$, none of which leads to a spatial deflection of the geodesics.
However, for vector fields and massive tensor fields, the traceless part of the metric perturbations lead to non-zero perturbed connections $\Gamma_{ij}^{(1)t} =  \Gamma_{tj}^{(1)i} = \partial_t h_{ij}/2$ in the flat-space limit, where $i,j\in (x,y)$, leading to both a spatial deflection and a time delay of null geodesics.}
Furthermore, the scalar cloud represented by Eq.\,(\ref{GAWF}) contains a $\sin \theta$ dependence in the wavefunction, suppressing its field value along geodesics propagating through the cloud that reach nearly face-on observers, such as in the case of the EHT observations of M87$^\star$~\cite{Akiyama:2019cqa} and SgrA$^\star$~\cite{EventHorizonTelescope:2022wkp,EventHorizonTelescope:2022wok,EventHorizonTelescope:2022exc}. Therefore in the main text we focus on vector and tensor clouds for which the signature we propose to study should be stronger.

Integrating both the radial and angular part of the energy density, we have the expressions for the total mass of a vector and tensor cloud 
\be 
M_{\rm cloud} = \frac{\pi \Psi_{0}^2 r_g}{\alpha^4}.\label{eq:MC}
\ee
Comparing Eq.\,(\ref{eq:MC}) with the BH mass $M_{\rm BH} = 8\pi m_{\rm pl}^2 r_g$ gives the mass ratio
\be\begin{split} \frac{ M_{\rm cloud}}{M_{\rm BH}} \approx \, 0.82\%\, \left(\frac{\Psi_{0}}{10^{17}\, {\rm GeV}}\right)^2\, \left(\frac{0.4}{\alpha}\right)^4
\approx \left\{\begin{array}{ll}
 5\times10^{-6}\, \left(\frac{\epsilon^T}{10^{-3}}\right)^2\, \left(\frac{1}{\alpha_h}\right)^2 \left(\frac{0.4}{\alpha}\right)^4\\
 1\times10^{-2}\, \left(\frac{\epsilon^V}{10^{-3}}\right)\, \left(\frac{0.4}{\alpha}\right)^4
\end{array}\right. , \label{eq:MC10}
\end{split}\ee
for the $(n\ell) = (10)$ state of a massive tensor and vector field, respectively. The mass for the $(2122)$ vector state differs from Eq.\,(\ref{eq:MC}) and Eq.\,(\ref{eq:MC10}) simply by a factor of $\approx 118$.

\begin{center}
\textbf{II: Photon ring autocorrelation}
\end{center}

The turbulent nature of accretion flows around black holes leads to significant fluctuations of the emission, characterized by a spatial and a temporal correlation length.
The correlation lengths can be inferred from general relativistic magnetohydrodynamic (GRMHD) simulations of accretion flow models that fit the EHT observations~\cite{Hadar:2020fda}.

To simulate the observations of a time-varying accretion flow, one needs to use backward ray tracing and covariant radiative transfer equations. The former method computes null geodesics starting from a local observer located at some far-away distance and at a fixed inclination angle with respect to the black-hole spin axis.
Using the black hole as the origin of the observer plane coordinate system, the distance to the origin gives the impact parameter of a given null geodesic.
Geodesics from different initial directions pass through accretion flows where the turbulent plasma emits photons and modifies their dispersion relation. 
These effects are taken into account by the process of covariant radiative transfer, which integrates quantities related to the photon intensity (the four independent Stokes parameters) along the geodesics.
Fluctuations in the emission then contribute to intensity fluctuations $\Delta I$ on the observer plane.

For a given emission source near the supermassive black hole, there are multiple geodesics connecting it and the observer plane, differing by the number of half orbits they do around the supermassive black hole.
The lensed photons reach the observer plane in a narrow ring region, locally enhancing the intensity and forming the photon ring observed.
Thus different points on the photon ring can come from the same emission sources. Taking into account the time delay due to lensing, the intensity fluctuations on these points share the same intensity fluctuations due to their common origin in the time-varying emission sources. Correlations between intensity fluctuations of points on the photon ring separated by a certain time-delay and azimuthal angle lapse are universally dependent on the space-time geometry in the photon shell. 

 \begin{figure}[htb]
  \includegraphics[width=0.4\textwidth]{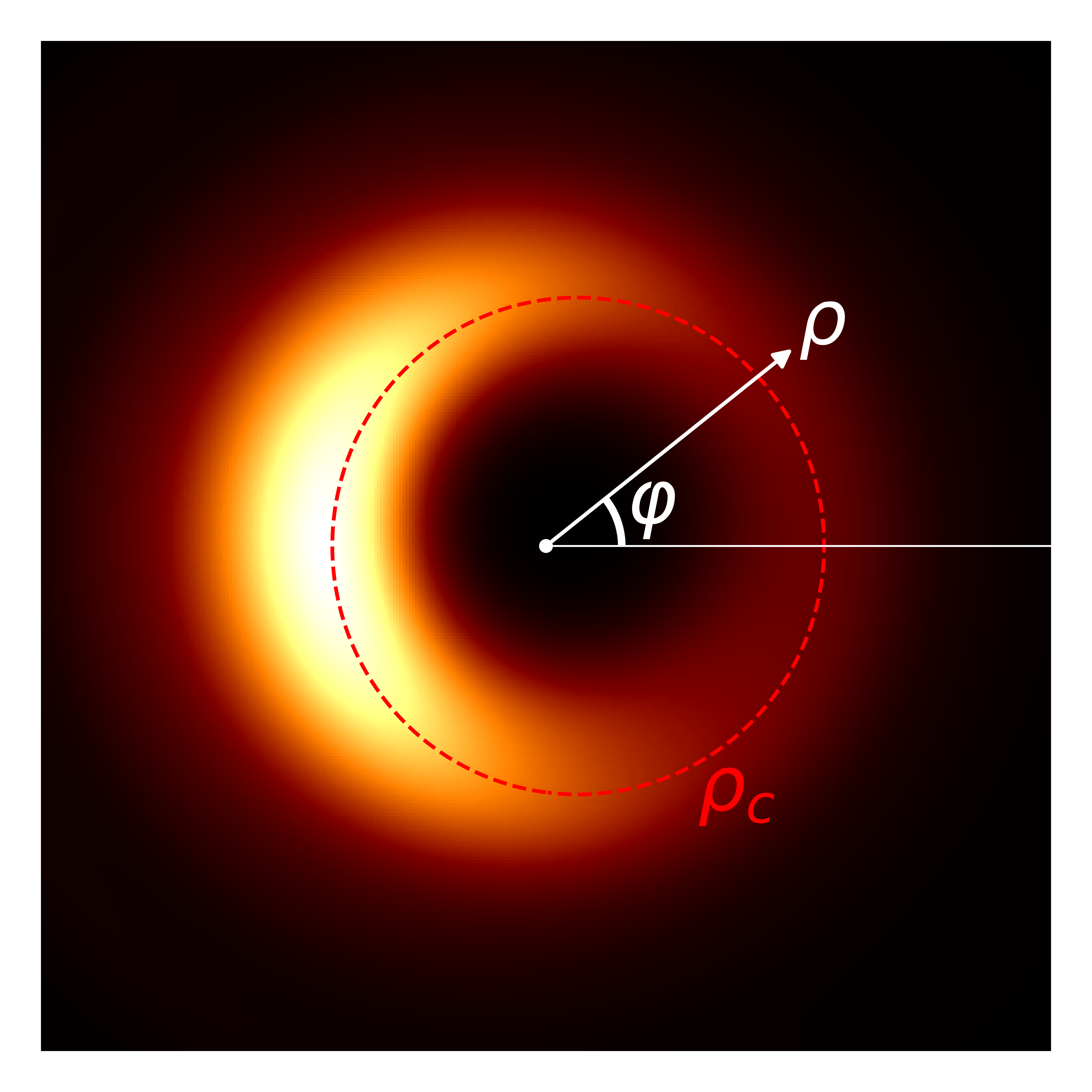}
 \includegraphics[width=0.52\textwidth]{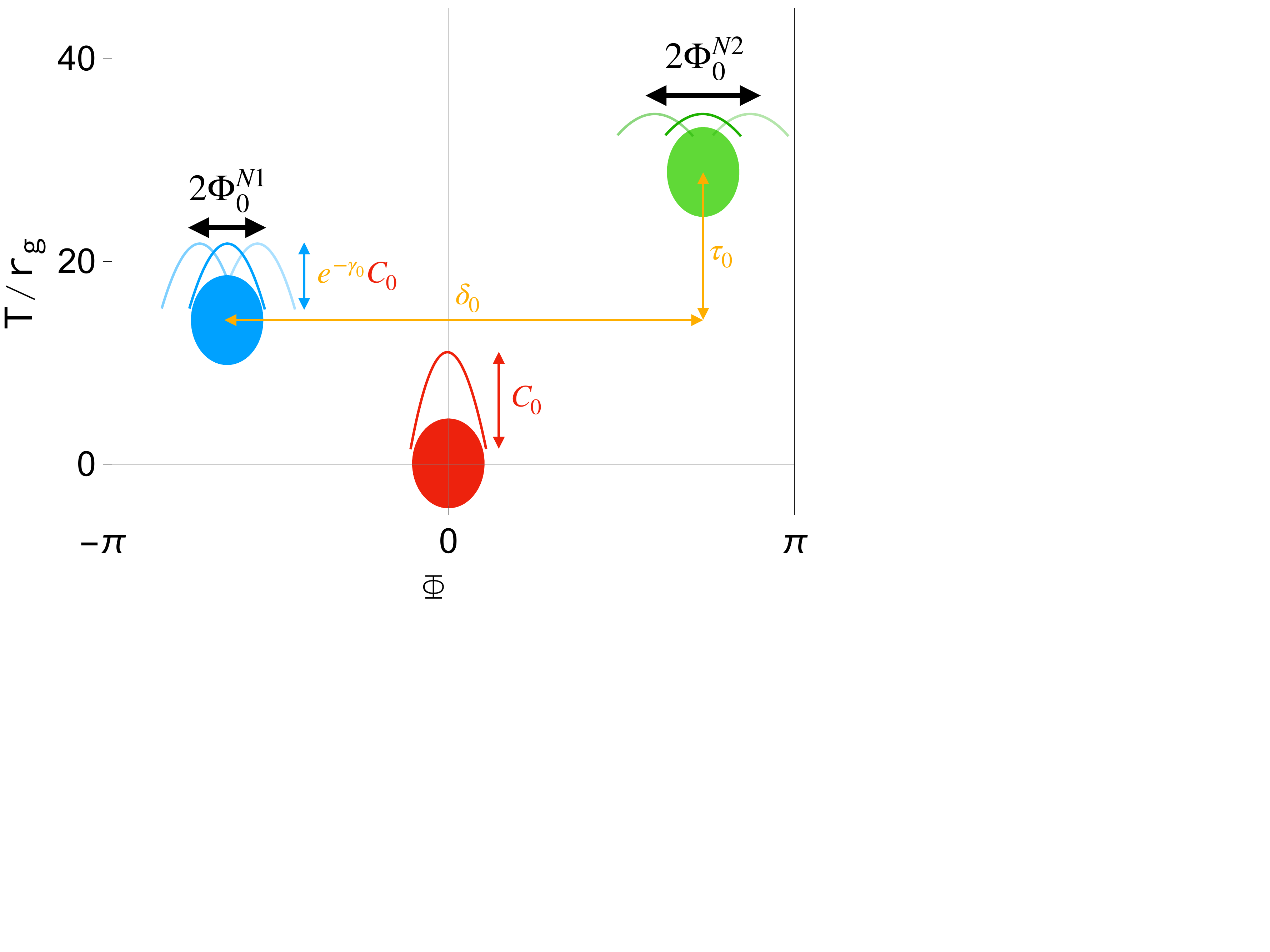}
     \caption{Left: image coordinate system $(\rho, \varphi)$ with the origin set at the black hole.
     The background intensity is calculated using a covariant radiative transfer code \texttt{IPOLE}~\cite{Moscibrodzka:2017lcu,Noble:2007zx},
     from an accretion flow outside the black hole with an inclination angle $\theta_{o} = 17^\circ$ and a BH dimensionless spin $a_J = 0.94$.
     The critical curve $\rho_{c}$ on the photon ring is shown in a red dashed line.
     Right: illustration of photon ring autocorrelations on the $(T, \Phi)$ plane and potential oscillations of the center of the peaks in the azimuthal direction, caused by the presence of a boson cloud.}
     \label{fig:illu}
 \end{figure}

An observable to quantify the correlations is the two-point correlation function computed on the photon ring~\cite{Hadar:2020fda}
\begin{equation}
    \label{eq:IFC}
    \mathcal{C}(T, \Phi)\equiv\iint \mathrm{d} \rho\, \mathrm{d} \rho^{\prime} \rho\,\rho^\prime \left\langle\Delta I(t, \rho, \varphi) \Delta I\left(t\!+\!T, \rho^{\prime}, \varphi\!+\!\Phi\right)\right\rangle,
\end{equation}
between intensity fluctuations $\Delta I$ convolved along the radial direction $\rho$, computed at times $t$ and $t+T$ and at an azimuthal angle coordinate in the ring $\varphi$ and $\varphi+\Phi$. For an optically thin accretion flow, one expects to see a series of peaks in the $(T, \Phi)$ plane representing the correlations, as shown by the colorful region in the right panel of Fig.\,\ref{fig:illu}. The red peak is the trivial self-correlation with value $\mathcal{C}_0$ being the local intensity, whose width is determined by the intrinsic correlation length of the emission source and a Gaussian convolution due to the finite observational spatial resolution. For a space-time dominated by a single black hole, other peaks are expected at $\{T \approx N \tau_0, \Phi \approx N \delta_0\}$, where $N \equiv k - k^\prime$ is the difference between the number of half orbits $k$ and $k^\prime$ each null geodesic of a given pair of intensity fluctuations executes around the black hole, and $\{\tau_0,\delta_0\}$ are the critical parameters characterizing the time delay and azimuthal lapse of the bound photon orbits.
The blue and green regions in Fig.\,\ref{fig:illu} are the correlations for $N=1$ and $N=2$, respectively. Their correlation strengths are suppressed by a factor of $e^{- N \gamma_0}$ compared to the one of the self-correlation due to a narrower region on the photon ring that can see lensed photons characterized by the critical parameter $\gamma_0 \approx \pi$~\cite{Gralla:2019drh}.

The signal-to-noise ratio (SNR) for the $N=1$ correlation peak is estimated to be $\textrm{SNR} \sim e^{-\gamma_0} \sqrt{N_{\rm eff}}$, 
where $N_{\rm eff}$ is the number of independent pair samples in the $(T, \Phi)$ plane after taking into account the finite spatial and temporal resolution~\cite{Hadar:2020fda}. Simple estimates based on the current EHT show that after one year of observation time the $N=1$ peak could be resolved~\cite{Hadar:2020fda}, while for the next-generation upgrade, ngEHT, the observation time needed to resolve the $N=1$ peak could be significantly shortened. The observation of the $N=2$ peak is more challenging, due to the further suppression in the correlation amplitude. Observing it requires longer observation times or a more coherent emitting source like a hotspot or a pulsar.
 
Finally, we give an estimate on the sensitivity to see the periodic oscillation of the correlation peak position, that would occur if a boson cloud is present, as predicted in the main text. Due to the finite spatial resolution contributing to a spatial Gaussian smearing, the correlation peak appears as a Gaussian packet on the $(T, \Phi)$ plane. For a spatial resolution  of $\sim 10 \mu$as as expected for ngEHT~\cite{Lico:2023mus}, the 1-$\sigma$ width of the Gaussian packet in the $\Phi$ direction is around $10^\circ$ for both M87$^\star$ and Sgr A$^\star$. A periodic oscillation of the central position of the Gaussian packet would broaden the width seen in the $\Phi$ domain, {as illustrated in the right panel of Fig.~\ref{fig:illu}}. {The broadening of the width becomes evident when the oscillation amplitude $\Phi_0^N$ is comparable to its width without oscillation.}
Thus a conservative constraint after a discovery of a correlation peak predicted by the general relativity is $2 \Phi_0^N < 10^\circ$. Notice that such a conservative estimate does not exploit the full oscillating features. One can check whether an oscillation in the center of the Gaussian packet gives a better likelihood than a stationary Gaussian packet, and leads to a sensitivity beyond the one limited by the Gaussian smearing. A longer observation time or a better spatial resolution can help resolve the intrinsic correlation of the plasma. {The main limitation in this case is the intrinsic correlation length in the azimuthal direction $\ell_\phi \approx 4.3^\circ$, obtained from GRMHD simulations for an accretion flow model that satisfies the EHT observations~\cite{Hadar:2020fda}. We therefore use $\ell_\phi$ as an optimistic limit to set our prospective constraints in the main text.}

\begin{center}
\textbf{III: Oscillation amplitude map on the photon ring}
\end{center}

\begin{figure}[htb]
 \includegraphics[width=0.78\textwidth]{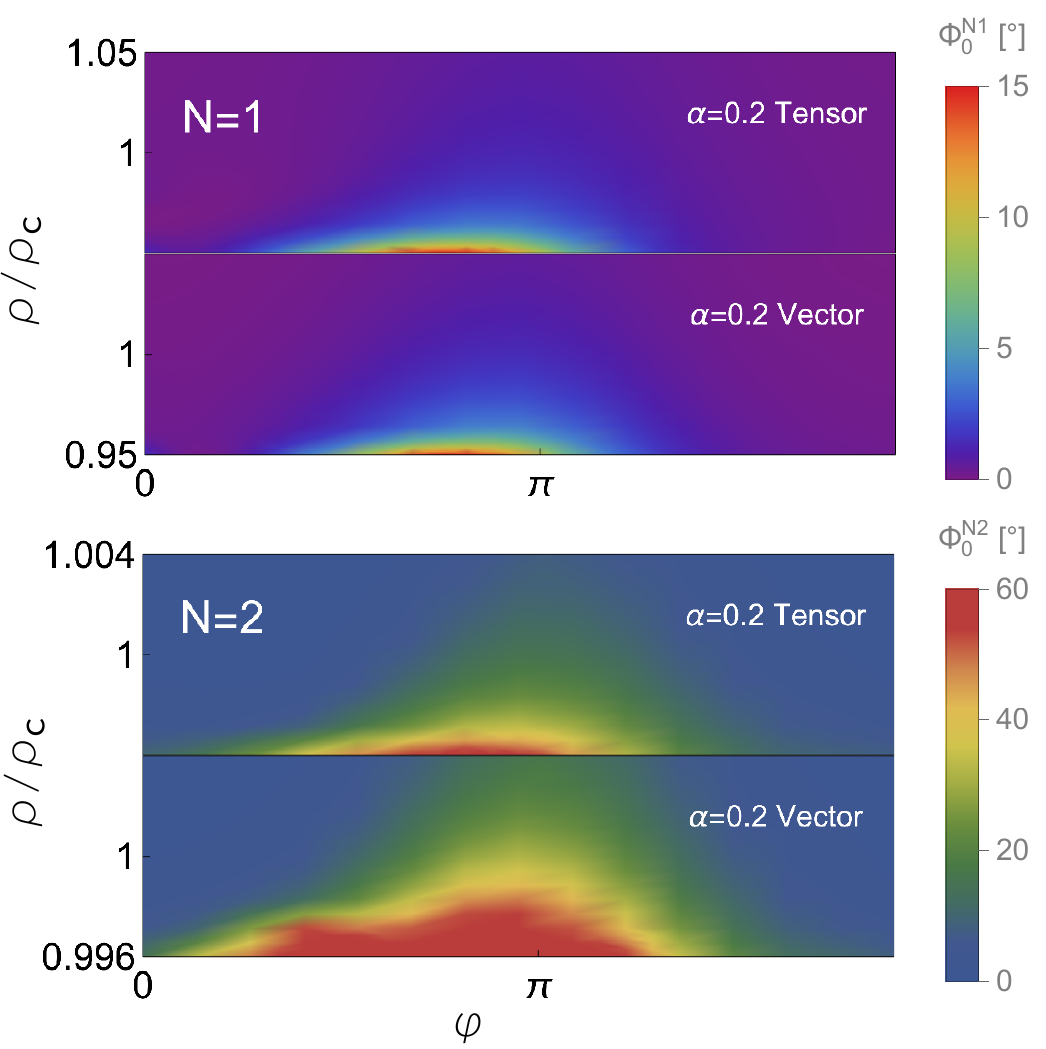}
     \caption{Oscillation amplitude $\Phi_0^N$ of the $N=1$ and $2$ azimuthal lapse as a function of the points on the image plane with $a_J = 0.94$ , $\theta_{o} = 17^\circ$, and $\epsilon = 10^{-3}$.
     The radial distance is measured in terms of the critical curve $\rho_{c} [\varphi]$.
     The $N=2$ case has a much narrower range in $\rho$ compared with $N=1$.}
     \label{fig:ampN1N2}
 \end{figure}

In Fig.\,\ref{fig:ampN1N2}, we show the map for $\Phi_0$ as a function of the image plane coordinates close to the photon ring for $\theta_{o} = 17^\circ$, a BH spin $a_J = 0.94$ and $\epsilon_T = \epsilon_V = 10^{-3}$, where $\varphi = \pi/2$ corresponds to the BH spin projection and $\rho_{c} (\varphi)$ is the radial distance between the critical curve and the BH. As expected, for $N=2$ with a much narrower range in $\rho$, $\Phi_0^{N2}$ is typically more significant due to the exponential sequential growth. Furthermore, in the inner region of the photon ring, corresponding to smaller values of $\rho/\rho_{c}$, $\Phi_0^N$ is larger, owing to a stronger instability of photon geodesics there and due to the fact that the azimuthal lapse deviation is inversely proportional to the radius of the emission source on the equatorial plane.

\begin{center}
\textbf{IV: Prospects for lower black hole spins}
\end{center}

\begin{figure*}[htb!]
\begin{tabular}{cc}
\includegraphics[width=0.45\textwidth]{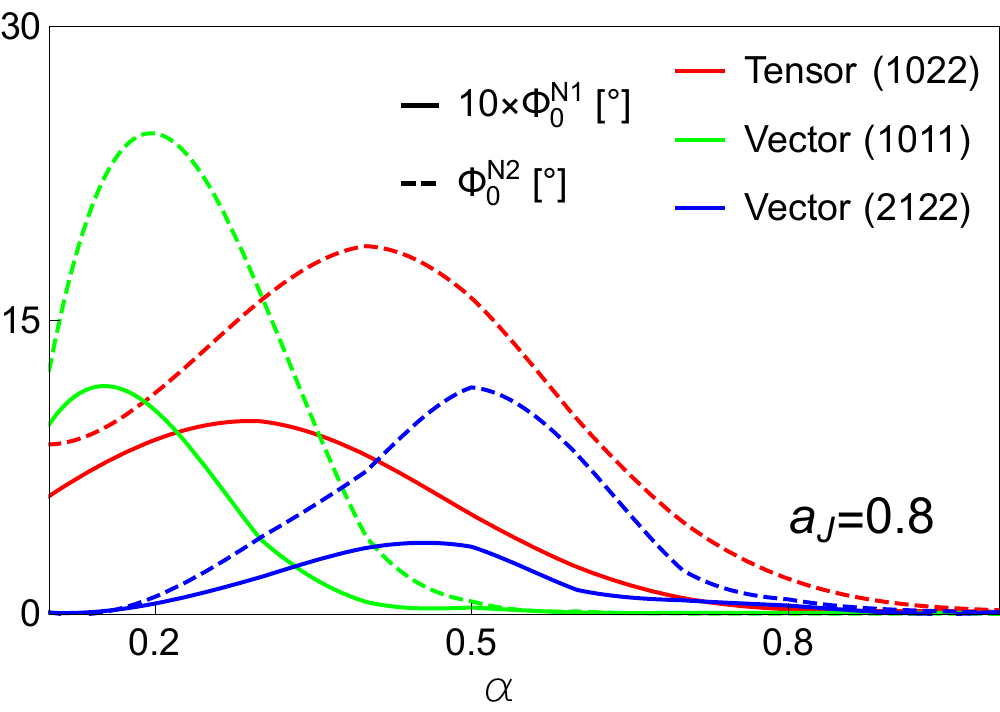} & \hspace{0.2cm}
 \includegraphics[width=0.45\textwidth]{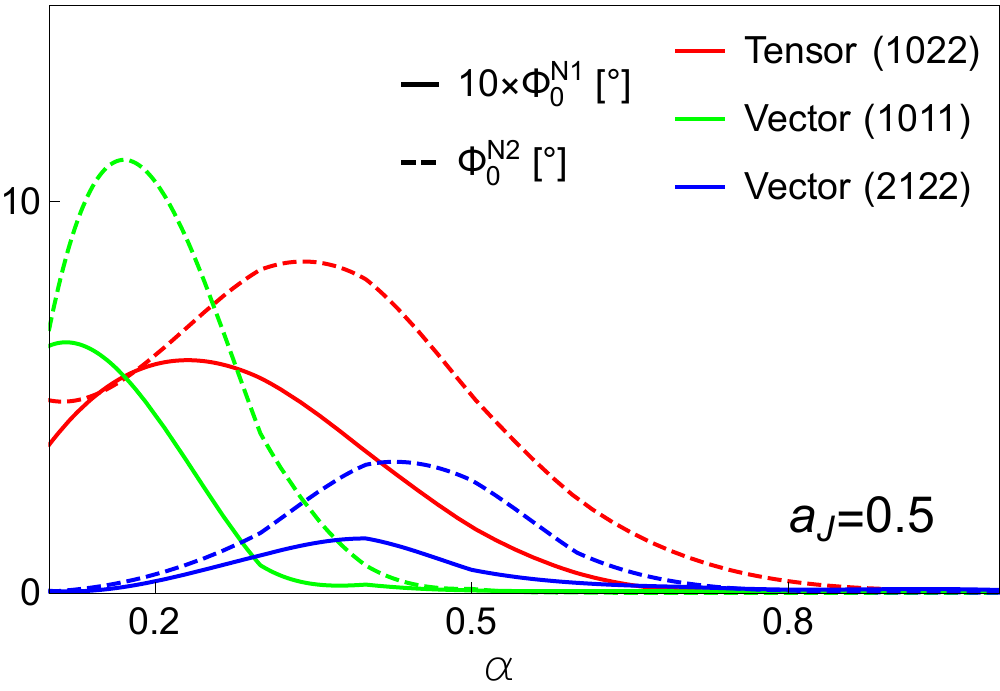} \\
 \includegraphics[width=0.45\textwidth]{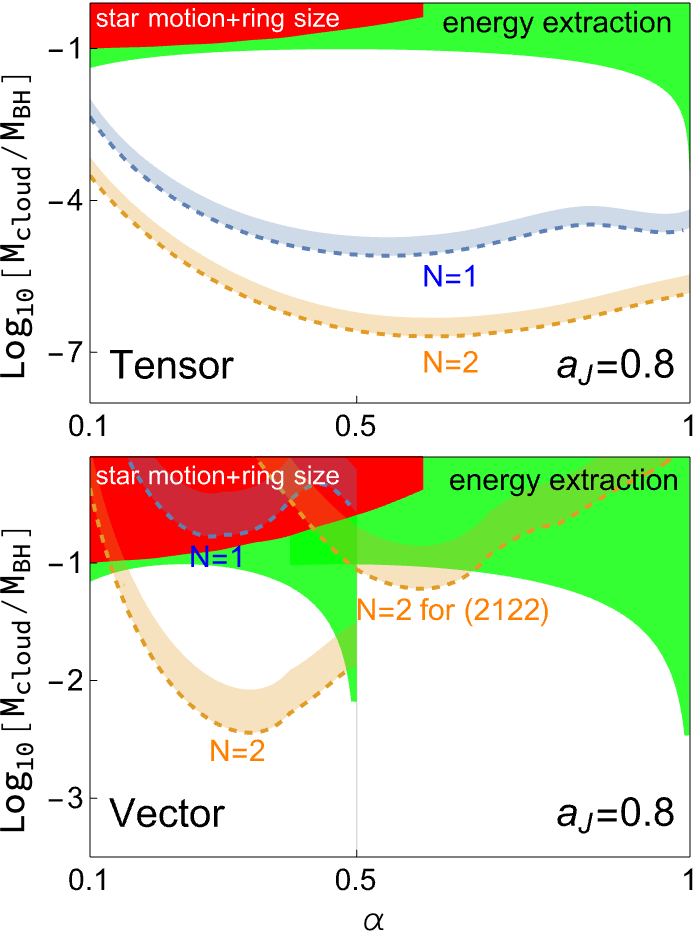} &
\includegraphics[width=0.45\textwidth]{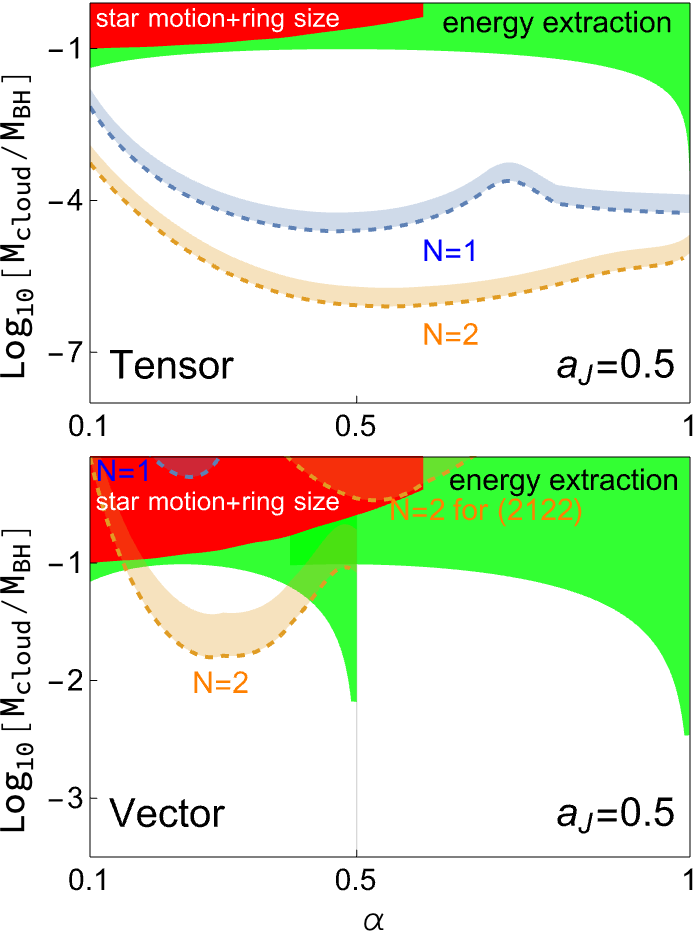}
\end{tabular}
    \caption{Same as Figs. 2 and 3 of the main text with $a_J=0.8$ (left column) and  $a_J=0.5$ (right column).}
       \label{fig:aJ08}
 \end{figure*}

In the main text, we showed benchmark calculations assuming a BH spin $a_J = 0.94$, motivated by previous claims that M87$^\star$  is a nearly extremal Kerr BH~\cite{Tamburini:2019vrf,Feng:2017vba} and the fact that EHT observations for Sgr\,A$^\star$ are consistent with large BH spins~\cite{EventHorizonTelescope:2022wkp,EventHorizonTelescope:2022urf}. On the other hand, modelling of M87$^\star$'s jet 
gives a more conservative bound with $a_J > 0.5$~\cite{Cruz-Osorio:2021}. In this appendix, we show results for lower spins, namely $a_J = 0.8$ and $0.5$, while keeping the other parameters as being the same as in main text. See Figs.\,\ref{fig:aJ08}. At low spin, the oscillations of the azimuthal lapse are typically smaller due to a larger photon ring orbit radius, which in turn implies slightly weaker constraints compared to $a_J=0.94$. However, even in the most pessimistic case of $a_J=0.5$, we find that the $N=2$ subring can always constrain interesting regions of the parameter space.

\end{document}